\documentclass{revtex4}

\begin{document}

\markboth{}
{}

\title{Gravitational energy of a Schwarzschild black hole}

\author{K. Shimizu}

\address{Division of Computer Science , The University of Aizu,
Aizu-Wakamatsu 965-8580, Japan}

\begin{abstract}
In a previous paper, we proposed a new gravitational energy momentum tensor. Here we use this tensor to evaluate the gravitational energies both inside and outside the horizon of a Schwarzschild black hole. Our results show that all of the gravitational energy exists outside the horizon, and that there is no gravitational energy inside the horizon. We comment on a relation with our
gravitational energy momentum tensor and another one which is proposed in a teleparallel gravity.
\end{abstract}

\maketitle

The definition of the gravitational energy momentum tensor (GEMT) has been a long-standing problem in the field of general relativity. Several definitions of the GEMT have been previously proposed. However, these are all different from each other, and are generally not proper tensors. \cite{landau}$,$\cite{moller}$,$ \cite{papapetrou}$,$\cite{weinberg}$,$\cite{padmanabhn}$,$\cite{misner}\par
We proposed a new GEMT in a previous paper. \cite{shimizu} It transforms correctly as a tensor under general coordinate transformations. One of the indices of the GEMT is a local Lorentz coordinate. The law of energy conservation is given by
\begin{equation}
\partial_\mu e(t^\mu_{\hspace{5pt}a}+T^\mu_{\hspace{5pt}a})=0.
\end{equation}
Here $t^\mu_a$ is the GEMT and $T^\mu_a$ is a matter energy momentum tensor. $e$ is the determinant of a tetrad. The Latin indices represent local Lorentz coordinates and the Greek indices represent world coordinates in 4-D space time. This equation is covariant under general coordinate transformations. General relativity is invariant under global Lorentz transformations. Therefore, the global Lorentz transformations do not break the law of energy conservation.\par
The GEMT of the Einstein-Hilbert action is given by
\begin{equation}
et(EH)^\mu_{\hspace{5pt}a}=\frac{1}{16\pi}(e^\mu_a e R-2eR^\mu_{\hspace{5pt}a}+\partial_\lambda e(c_a^{\hspace{5pt}\lambda\mu}-c^{\lambda\mu}_{\hspace{10pt}a}+c^{\mu\lambda}_{\hspace{10pt}a})),\label {emt}
\end {equation}
where $c_a^{\hspace{5pt}\lambda\mu} \equiv g^{\lambda \sigma}g^{\mu\nu}(e_{a\sigma,\nu}-e_{a\nu,\sigma})$ and $c^{\lambda\mu}_{\hspace{10pt}a}\equiv e^\lambda_b g^{\mu\nu}e^\rho_a(e^b_{\nu,\rho}-e^b_{\rho,\nu})$. We take units c=G=1. If we write the third term covariantly, an anti-symmetric term appears. However, because a relation $\nabla_\mu c^{\mu a b}=0$ must be satisfied by the law of conservation of the spin angular momentum tensor, the GEMT is symmetric.\par
When we consider a restricted region of space time, we add the GEMT of the Gibbons-Hawking term and its subtraction term. These are given by
\begin{equation}
t^i_\alpha(GH)= -\frac{1}{8\pi}\frac{1}{\sqrt{|g^{\hspace{1pt}\zeta\zeta}|}}(e^i_\alpha c^{\lambda\zeta}_{\hspace{10pt}\lambda} +\frac{\partial{c^{\lambda\zeta}_{\hspace{10pt}\lambda}}}{\partial{e^\beta_{k,i}}}c^\beta_{\hspace{5pt}\alpha k} ),\label{GH}
\end {equation}
and
\begin{equation}
t^i_\alpha(GH)_0= -\frac{1}{8\pi}\frac{1}{\sqrt{|\bar{g}^{\hspace{1pt}\zeta\zeta}|}}(e^i_\alpha \bar{c}^{\lambda\zeta}_{\hspace{10pt}\lambda} +\frac{\partial{\bar{c}^{\lambda\zeta}_{\hspace{10pt}\lambda}}}{\partial{e^\beta_{k,i}}}c^\beta_{\hspace{5pt}\alpha k})).\label{GH0}
\end {equation}
We assume that the world coordinate $\zeta=const.$ hyper surface represents a boundary and that one of the tetrads is perpendicular to the surfaces of this boundary. $i$ and $k$ are world coordinates, $\alpha$, and $\beta$ are local Lorentz coordinates of the 3-D space time of the boundary.
$\lambda$ is a world coordinate and $a$ is a local Lorentz coordinate in 4-D space time. $\bar{c}^{\lambda\zeta}_{\hspace{10pt}\lambda}$ is a flat space time quantity.\par

In this article, we evaluate the gravitational energy of a Schwarzschild black hole.\cite{castello}
The metric of the Schwarzschild black hole is
\begin{equation}
ds^2=-\Bigl(1-\frac{2M}{r}\Bigr)dt^2+\Bigl(1-\frac{2M}{r}\Bigr)^{-1}dr^2+r^2d\Omega^2.
\end{equation}
We first consider the energy outside the horizon. The energy density is given by
\begin{equation}
\varepsilon=t_{\mu\nu}n^\mu n^\nu,
\end{equation}
where $n^\mu$ is the 4-velocity of an observer. Here, we assume a static observer. The 4-velocity at radius r is
\begin{equation}
n^\mu=\bigl(\frac{1}{\sqrt{1-2M/r}},0,0,0 \bigr).
\end{equation}
Because $n^\mu$ only has a time component, we evaluate only the $t_{t t}$ component of the GEMT.\par

The coordinate system of the static observer determines the tetrad.\cite{castello}$,$ \cite{maluf}$,$\cite{maluf3} It is given by
\[e_{a\mu}=\left(
           \begin{array}{cccc}
            -\frac{1}{A} & 0 & 0 & 0 \\
            0 & Asin\theta cos\phi & rcos\theta cos\phi & -rsin\theta sin\phi \\
            0 & Asin\theta sin\phi & rcos\theta sin\phi &  rsin\theta cos\phi \\
            0 & -Acos\theta & -rsin\theta & 0
           \end{array}
          \right),  \]
where $A\equiv\frac{1}{\sqrt{1-2M/r}}$. A naive tetrad could not be defined at the origin of coordinate.

Because the Schwarzschild black hole is a vacuum solution of the Einstein field equations , the $t^t_0$ component of the GEMT of the Einstein-Hilbert action is
\begin{eqnarray}
t(EH)^t_0&=&\frac{1}{16\pi e}\partial_\lambda e(c^{t\lambda}_{\hspace{10pt}0}-c^{\lambda t}_{\hspace{10pt}0}+c_0^{\hspace{5pt}\lambda t}) \nonumber \\
&=&\frac{1}{8\pi e}\partial_r ec_0^{\hspace{5pt}r t} \nonumber \\
  &=&\frac{M^2}{8\pi r^4 (1-2M/r)^{3/2}}.
\end{eqnarray}
The gravitational energy of the Einstein-Hilbert action from $r=2M$ to $r=R$ is
\begin{eqnarray}
E(EH)&=&\int^R_{2M} d^3x\sqrt{\gamma}t_{tt}n^t n^t \nonumber \\
     &=&\int^R_{2M} d^3x\sqrt{\gamma}t_0^t g_{tt} e^0_t n^t n^t \nonumber \\
     &=&\frac{M}{2\sqrt{1-2M/R}}-\frac{M}{2\sqrt{0}}.
\end{eqnarray}
Next, we consider the GEMT of the Gibbons-Hawking term. The $t^t_0$ component of the GEMT is

\begin{eqnarray}
t^t_0(GH)&=&-\frac{1}{8\pi}e^t_0\frac{1}{\sqrt{g^{rr}}}\left(c^{\lambda r}_{\hspace{10pt}\lambda}
+\frac{\partial{c^{\lambda r}_{\hspace{10pt}\lambda}}}{\partial e^{\beta}_{k,t}}c^{\beta}_{\hspace{5pt}0 k}\right) \nonumber \\
         &=&\frac{1}{8\pi}\left(\frac{M/r^2}{1-2M/r}-\frac{2}{r\sqrt{1-2M/r}}+\frac{2}{r}\right).
\end{eqnarray}
At $r=R$, the gravitational energy of the Gibbons-Hawking term becomes
\begin{eqnarray}
E(GH)_{r=R}&=&\int d^2x\sqrt{\sigma}t_{tt}(GH)n^t n^t|_{r=R} \nonumber \\
           &=&-\frac{M}{2\sqrt{1-2M/R}}+M.
\end{eqnarray}
At $r=2M$, the gravitational energy of the Gibbons-Hawking term becomes
\begin{equation}
E(GH)_{r=2M}=-2M+\frac{M}{2\sqrt{0}},
\end{equation}
as the trace of the extrinsic curvature is $K=\nabla_\mu n^\mu$, where $n^\mu$ is the outwards-pointing normal vector.  So at $r=2M$, $n^\mu$ points towards the origin of the coordinate system. Its direction is the inverse of that at $r=R$. Therefore, the signature of the energy is inverse to that at $r=R$. The $t^t_0$ component of the GEMT of the subtraction term is zero. \par
Therefore, the total gravitational energy outside the horizon is
\begin{eqnarray}
E_{out}&\equiv& \lim_{R \to \infty}(E(EH)+E(GH)_{r=R}+E(GH)_{r=2M}) \nonumber \\
       &=&-M.
\end{eqnarray}
\par
Next, we consider the gravitational energy inside the horizon. Within the event horizon, the r-axis plays the role of time.
Therefore, if we consider an observer who moves along a $t=const.$ line, the observer is static. Then the 4-velocity
of the observer is $ u^\mu=(0,\sqrt{2M/r-1},0,0)$.\par
In the same way as that used to derive the tetrad outside the horizon, we obtain the tetrad inside the horizon as follows:
\[e_{a\mu}=\left(
           \begin{array}{cccc}
            Bsin\theta cos\phi & 0             & rcos\theta cos\phi & -r sin\theta sin\phi \\
            0                   & -\frac{1}{B}  & 0                  & 0                    \\
            Bsin\theta sin\phi & 0             & rcos\theta sin\phi & rsin\theta cos\phi \\
            Bcos\theta         & 0             & -rsin\theta         & 0
           \end{array}
          \right), \]
where $B=\sqrt{2M/r-1}$.    \par

We evaluate the GEMT of the Einstein-Hilbert action. Because only $u^r$ is non-zero, we only need $t(EH)^r_1$.
This can be obtained as
\begin{eqnarray}
t(EH)^r_1&=&-\frac{1}{16\pi e}\partial_\lambda e(c^{r\lambda}_{\hspace{10pt}1}-c^{\lambda r}_{\hspace{10pt}1}+c_1^{\hspace{5pt}\lambda r}) \nonumber \\
         &=&0.
\end{eqnarray}
Because the boundary is the $r=2M$ hyper surface, the $t(GH)^r_1$ component of the GEMT is not defined. Therefore, the gravitational energy
inside the horizon is zero. The gravitational energy of the Schwarzschild black hole exists entirely outside its horizon and its
value is the negative of the black hole mass. \par

In a previous paper {\it Ref.} 10, the authors evaluated the gravitational energy of a charged black hole. In their case, the gravitational energy enclosed by a spherical surface of constant radius r was given by $ E(r)=r(1-\sqrt{1-2M/r+Q^2/r})$. This is clearly different from our result, as the gravitational energy is positive. Some researchers of teleparallel gravity have also proposed an alternative GEMT,\cite{maluf2}$,$ \cite{maluf4} which is also different from ours. It is given by

\begin{equation}
16\pi t^\mu_a=c^{\sigma\lambda}_{\hspace{10pt}a}(c_{\sigma\lambda}^{\hspace{10pt}\mu}+c_{\lambda\sigma}^{\hspace{10pt}\mu}-c^\mu_{\hspace{5pt}\sigma\lambda})-2c^\sigma_{\hspace{5pt}a \sigma}c^{\lambda\mu}_{\hspace{10pt}\lambda}-2c^{\sigma\lambda}_{\hspace{10pt}\sigma}c^\mu_{\hspace{5pt}\lambda a}+e^\mu_a(R-2\nabla_\sigma c^{\lambda\sigma}_{\hspace{10pt}\lambda})
\end{equation}
in our notation. We can rewrite this GEMT as follows:
\begin{eqnarray}
 t^\mu_a&=&t^\mu_a(1)+t^\mu_a(2),\nonumber \\
 16\pi t^\mu_a(1)&\equiv&e^\mu_a R-2R^\mu_a+\nabla_\lambda(c^{\mu\lambda}_{\hspace{10pt}a}+c_a^{\hspace{5pt}\lambda\mu})
-\frac{1}{2}c_{a\lambda\sigma}c^{\mu\lambda\sigma}+c_{\lambda\sigma a}(c^{\lambda\sigma\mu}+c^{\sigma\lambda\mu}),\\
16\pi t^\mu_a(2)&\equiv&\nabla_a c^{\lambda\mu}_{\hspace{10pt}\lambda}+\nabla^\mu c^{\lambda}_{\hspace{5pt}a\lambda}-2e^\mu_a\nabla_\sigma c^{\lambda\sigma}_{\hspace{10pt}\lambda}-2c^{\lambda}_{\hspace{5pt}a\lambda}c^{\sigma\mu}_{\hspace{10pt}\sigma} -c^{\sigma\lambda}_{\hspace{10pt}\sigma}(c^\mu_{\lambda a}-c_{a\lambda}^{\hspace{10pt}\mu})\nonumber \\
 & &{}-\frac{1}{2}(c^{\mu\sigma\lambda}c_{\sigma\lambda a}-c_{a\lambda\sigma}c^{\lambda\sigma\mu}).
\end{eqnarray}
Because $ \partial_\mu \,e \,t^\mu_a(2)=0$ is satisfied, the energy conservation law becomes
\begin{eqnarray}
  & & \partial_\mu e(t^\mu_a(1)+t^\mu_a(2)+T^\mu_a) \nonumber \\
  &=& \partial_\mu e(t^\mu_a(1)+T^\mu_a) \nonumber \\
  &=& 0.
\end{eqnarray}
We should therefore consider $t^\mu_a(1)$ to be the GEMT, which is equivalent to our own GEMT.

\section*{Acknowledgments}
The author would like to sincerely thank Professor Akira Fujitsu contributing to discussions that were helpful in the production of this article.


\section*{References}

\vspace*{6pt}

\begin{thebibliography}{99}
\bibitem{landau} L. D. Landau and E. M. Lifshitz, {\it the classical Theory of Fields}Pergamon Press, Neew York, (1975)
\bibitem{moller} C. M\o ller, Ann.  Phys. {\bf 4}, 347 (1958)
\bibitem{papapetrou} A.~Papapetrou, {\it Proc.Roy.Irish Acad.(Sect.A)} {\bf 52A}, 11 (1948)
\bibitem{weinberg}  S.~Weinberg, {\it Gravitation and Cosmology} (Wiley, New York, 1972)
\bibitem{padmanabhn} T.~Padmanabhn, {\it Gravitation} (Cambridge Univ. Press, Cambridge, 2010)
\bibitem{misner} C.~W.~Misner, K.~S.~Thorne and J.~A.~Wheeler, {\it Gravitation} (W.~H.~Freeman and Company, 1973)
\bibitem{shimizu} K. Shimizu, {\it  Mod. Phys. Lett. A} {\bf 31}, 1650151 (2016) 10.1142/S0217732316501510
\bibitem{castello} K. H. C. Castello-Branco and J. F. da Rocha-Neto arXiv:1311.1590
\bibitem{maluf} J. W.Maluf, F. F. Faria and S. C. Ulhoa, {\it class. Quant. Grav.} {\bf 24}, 2743 (2007)
\bibitem{maluf3} J. W.Maluf, J. F. da Rocha-Neto, T. M. L. Toribio and K. H. Castello-Branco, {\it Phys. Rev.} {\bf D65}, 124001 (2002)
\bibitem{castello2} K. H. C. Castello-Branco and J. F. da Rocha-Neto, {\it Phys. Rev} {\bf D88}, 024045 (2013)
\bibitem{maluf2} J. W. Maluf, {\it Ann. Phys.} {\bf 525} 339 (2013)
\bibitem{maluf4} J. W. Maluf and S. C. Ulohoa, {\it Phys. Rev} {\bf D78}, 069901 (2008)

\end{thebibliography}
\end{document}